\documentclass[%
 reprint,
superscriptaddress,
nofootinbib,
 amsmath,amssymb,
 aps,
pra,
twocolumn,
]{revtex4-2}

\usepackage{graphicx}
\usepackage{dcolumn}
\usepackage{bm}
\usepackage{float}
\usepackage{graphicx}
\usepackage{amsmath}
\usepackage{textcomp}
\usepackage{tabularx}
\usepackage{calrsfs}
\DeclareMathAlphabet{\pazocal}{OMS}{zplm}{m}{n}

\usepackage{siunitx}
\usepackage{xcolor}
\usepackage{svg}
\newcommand{\kcb}[1]{\textcolor{black}{#1}}
\usepackage{comment}
\usepackage{amssymb}
\usepackage[hidelinks]{hyperref}
\usepackage{soul}
\usepackage{ulem}
\usepackage{url}
\usepackage{hyperref}
\hypersetup{
  breaklinks=true,
}

\begin{document}

\preprint{APS/123-QED}

\title{Phonon trapping lateral field excited suspended \kcb{overtone} bulk acoustic wave resonators (XBARs) \kcb{as a platform for microwave to optical signal transduction} }

\author{Elnaz Shokati}
\affiliation{Quantum Engineering Technology Labs and School of Electrical, Electronic and Mechanical Engineering, University of Bristol, Woodland Road, Bristol, UK, BS8 1UB}

\author{Robert Thomas}
\affiliation{Quantum Engineering Technology Labs and School of Electrical, Electronic and Mechanical Engineering, University of Bristol, Woodland Road, Bristol, UK, BS8 1UB}

\author{Krishna C. Balram}
\email{krishna.coimbatorebalram@bristol.ac.uk}
\affiliation{Quantum Engineering Technology Labs and School of Electrical, Electronic and Mechanical Engineering, University of Bristol, Woodland Road, Bristol, UK, BS8 1UB}

\begin{abstract}
Film bulk acoustic wave resonators (FBARs) underpin modern wireless communication by enabling compact, high-performance RF filters in modern smartphones. Traditionally, these FBAR devices work with quasi-plane waves of sound where the transverse extent of the acoustic field $\gg$ the acoustic wavelength ($\lambda_a$). On the other hand, strong modal confinement is needed for achieving the interaction strengths necessary for building efficient microwave to optical \kcb{signal} transducers (MW-OT) around an FBAR opto-mechanical cavity platform. \kcb{While MW-OTs have traditionally been engineered around sub-\qty{}{\um} scale optomechanical cavities, bulk acoustic wave approaches have inherent advantages in phonon injection efficiency, optical power handling and manufacturability. A key limitation of the FBAR geometry is that the acoustic field is confined under the metal electrode which makes it challenging to engineer the small mode-volume, high quality factor optical cavities which are critical for achieving high transduction efficiency.} Here, we \kcb{consider lateral field excited suspended overtone bulk acoustic wave resonators (XBARs) as an alternative bulk wave platform for MW-OT, which overcome this limitation, and outline the requirements needed for building efficient MW-OT around this geometry. As a first step towards viability, we} fabricate a small mode-volume phonon trapping acoustic microresonators by shaping the piezoelectric layer into a spherical lens and show an improvement in modal confinement and quality factor ($\approx$ 4$\times$). 

\end{abstract}

%\keywords{Suggested keywords}%Use showkeys class option if keyword
                              %display desired

\maketitle

\section{\label{sec:level1}Introduction}

Resonant piezoelectric MEMS devices \cite{hashimoto2009rf, bhugra2017piezoelectric}, especially film bulk acoustic wave resonators (FBARs), underpin modern wireless communication by being the de-facto standard for implementing compact, high-performance multiband RF filters in modern smartphones \cite{ruby2015snapshot}. Especially at high-frequencies ($>$ \qty{1}{\giga\hertz}), FBARs provide the necessary combination of high electro-mechanical coupling strengths ($k^2_{eff}$) and mechanical quality factor ($Q_m$) that enables one to meet the stringent device requirements in terms of bandwidth, insertion loss and frequency roll-off that modern filters require, especially as we move beyond the 5G-era \cite{mahon20175g, hagelauer2022microwave}. Despite their spectacular success and continued optimization \cite{ruby2011decade}, one thing has remained relatively constant across the many device iterations: FBAR devices manipulate quasi-plane waves of sound \cite{kc2024piezoelectric}. The transverse extent of the acoustic field is $\gg$ the acoustic wavelength ($\lambda_a$). Historically, this is because resonator size is primarily set by impedance (\qty{50}{\ohm}) matching constraints and there has been no real device-level driver for pushing strong acoustic field confinement.

On the other hand, there are other applications where strong field confinement provides a significant advantage. The prototypical example is a piezoelectric microwave to optical signal transducer (MW-OT) \cite{balram2022piezoelectric}. These devices are critical for networking remote superconducting qubit processors \cite{bardin2020quantum} via optical fiber networks \cite{bravyi2022future}, with a view towards achieving the scale necessary for achieving quantum advantage \cite{hoefler2023disentangling} on tasks such as factoring and computational chemistry. In an MW-OT, the microwave photon from a superconducting qubit is first converted to a mechanical mode and the optical photon conversion is mediated by an acousto-optic (AO) interaction in an opto-mechanical cavity \cite{aspelmeyer2014cavity} which supports resonant optical and mechanical modes with strong AO interaction.

To see why strong mechanical mode confinement is essential, it is helpful to consider a simplified AO interaction occurring in a 1D Fabry-Perot like cavity, where the AO interaction can be approximated \cite{renninger2018bulk} as:
\begin{equation}
    g\approx\frac{\omega^2_cn^3p_{12}}{2c}\sqrt{\frac{\hbar}{2{\rho}A_{eff}L\Omega_m}}
    \label{eqn:SBS_g}
\end{equation}

In this expression the AO interaction strength ($g$) [\qty{}{\radian\per\second}] depends on a number of parameters such as the optical frequency ($\omega_c$), material refractive index ($n$), photoelastic coefficient ($p_{12}$) mediating the interaction, density ($\rho$), speed of light ($c$) and Planck's constant ($\hbar$) which are usually fixed once we pick a material platform of interest. The mechanical frequency ($\Omega_m$) is usually set by the need to achieve phase matching (related to the Brillouin scattering frequency for bulk materials) and compatibility with current superconducting qubit operation frequencies (5-8 \qty{}{\GHz}). Given these constraints, the only way to maximize $g$ is to reduce the cavity mode-volume ($A_{eff}L$), where $A_{eff}$ is the transverse acoustic field extent (assumed identical to the optical beam to maximize $g$) and $L$ is the cavity length. Therefore, to maximize $g$ in the pursuit of high photon conversion efficiency (${\propto}g^2$), one needs strong acoustic modal confinement.

This general argument underlies why many of the state-of-the-art piezoelectric MW-OTs \cite{mirhosseini2020superconducting, jiang2023optically, weaver2024integrated} work around small mode volume integrated photonic devices, usually 1D opto-mechanical crystals which support a breathing mode with strong AO interaction \cite{chan2012optimized}. While they have continuously advanced in photon conversion efficiency ($\approx 5\%$ presently) \cite{jiang2023optically}, they face some fundamental issues which make it worth considering alternative transducer architectures. In particular, the 1D photonic crystal geometry, and its inherent $\lambda_a$-scale mechanical mode confinement make it challenging to excite the mechanical mode efficiently via standard piezoelectric transducers \cite{balram2022piezoelectric}. One can circumvent this problem to an extent by using mechanical mode hybridization \cite{wu2020microwave}, but realizing these doubly resonant structures while preserving high $g$ increases the fabrication complexity significantly \cite{weaver2024integrated}. \kcb{As a case in point, current MW-OT designs are unable to leverage established foundry infrastructure which limits both scalability and adoption.} A second issue is the large surface-to-volume ratio in these cavities which, via surface absorption, limits the optical pump power that can be used at \qty{}{\milli\kelvin} temperatures, thereby putting a bound on the achievable efficiency.

As an alternative architecture for building efficient MW-OT\cite{balram2022piezoelectric}, \kcb{bulk acoustic waves can be utilized in a doubly resonant configuration wherein both optical and acoustic fields are confined in a small mode-volume cavity. This approach has inherent advantages in fabrication simplicity and phonon injection efficiency by building on established piezoelectric resonator designs. The main tradeoff is a weaker AO interaction strength $g$, but the higher optical pump powers in the cavity can enhance the nonlinear interaction and partially compensate for the $g$ reduction.} 

High-overtone bulk acoustic wave resonator (HBAR) devices have demonstrated extremely low acoustic dissipation at cryogenic temperatures \cite{galliou2013extremely, gokhale2020epitaxial} and can be efficiently interfaced with superconducting qubits \cite{chu2017quantum} and light \cite{kharel2019high, doeleman2023brillouin, diamandi2025optomechanical}. These elements can be combined to build a bulk MW-OT \cite{yoon2023simultaneous} around the HBAR geometry but the overall transduction efficiency is very weak ($\approx 10^{-8}$). This is predominantly due to the reduction in $g$ via the mode volume scaling argument outlined above. For reference, $g/{2\pi}$ [\qty{}{\hertz}] in bulk systems is $\approx$ \qty{5}{\hertz} \cite{yoon2023simultaneous} whereas it is $\approx$ \qty{1}{\mega\hertz} in nanobeam cavities \cite{balram2014moving}. A second issue is that the traditional HBAR geometry \cite{hashimoto2009rf} is not well-suited for optical interfacing because the acoustic field is confined directly under the \kcb{metal} electrode, which means that an optical cavity built around this geometry \cite{valle2019high} needs to use the metal as (at least) one of the cavity mirrors. \kcb{This} reduces the optical quality factor \kcb{($Q_o$)} and has a knock-on effect on conversion efficiency. \kcb{A final issue is that high acoustic velocity of the longitudinal waves involved in HBAR devices pushes the interaction frequency beyond the preferred operating frequency range (5-8 \qty{}{\GHz}) \cite{bardin2020quantum} of current transmon qubits.}

In this work, we take a first step towards addressing these problems. To reduce the acoustic cavity mode volume and increase $g$, we extend the phonon trapping approach \cite{galliou2013extremely, kharel2018ultra} from \qty{}{\milli\meter}-thick ($L$) HBAR devices to demonstrate confocal cavities in suspended \qty{}{\um}-thick membranes. In addition, to facilitate efficient optical interfaces, \kcb{especially the small mode-volume, high optical quality factor ($Q$) cavities \cite{hunger2010fiber} needed to maximize $g$,} we move away from traditional thickness field excited HBAR and FBAR modes towards lateral field actuated XBAR modes \cite{koulakis2021xbar,yandrapalli2022study}. In such XBAR devices, the acoustic field is trapped in between the metal electrodes and they provide a natural route towards electrical actuation of the mechanics, without interfering with the optical path. We show that these two ideas can be combined to demonstrate high-frequency small mode volume acoustic cavities with reduced dissipation. While we focus primarily on the acoustics aspects in this work, the need to incorporate optics around this device architecture imposes certain tradeoffs inherent in this work that can be relaxed if these ideas are applied to purely microwave devices \cite{yandrapalli2022study}.

\section{Design requirement for XBAR overtone based MW-OTs}

\begin{figure}[htbp]
\centering
\includegraphics[width=1\columnwidth]{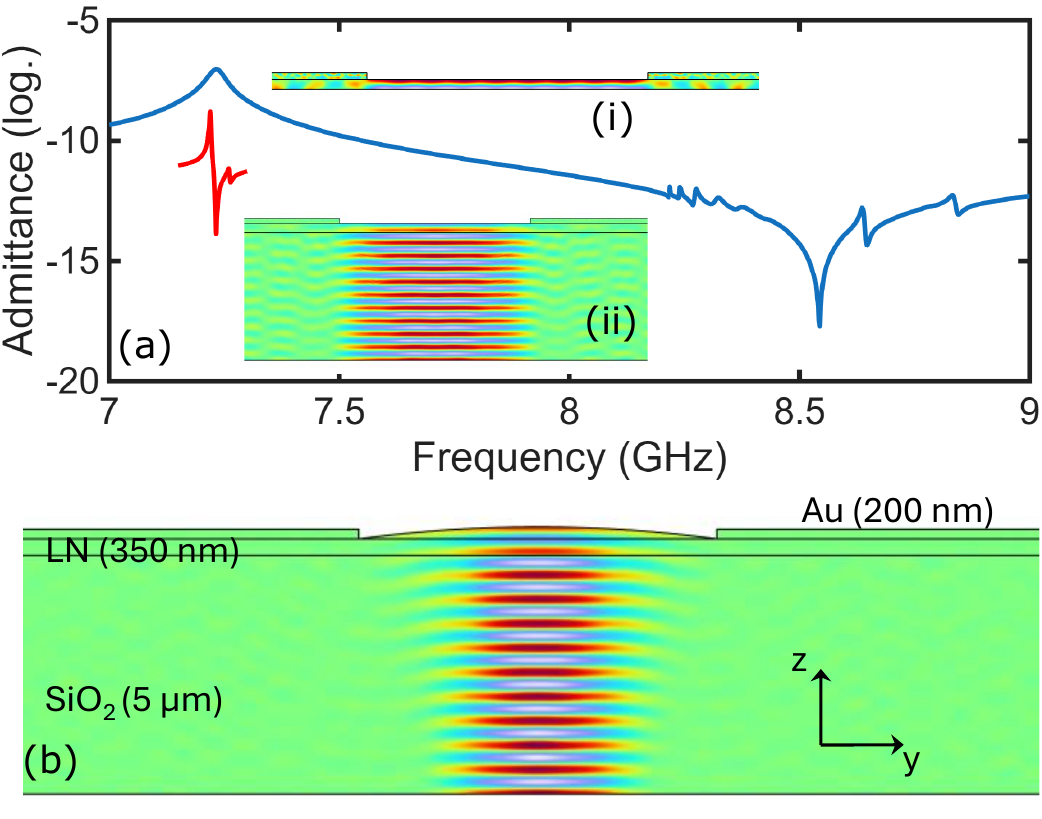}\\
\caption{Lateral field excited bulk acoustic wave resonators (XBAR) in thin film lithium niobate: (a) Admittance [\qty{}{dB}] of XBAR devices extracted from a 2D FEM simulation with out of plane dimension (thickness) set to \qty{20}{\um}. Blue curves shows the standard XBAR resonance ($A_1$ Lamb) in a \qty{250}{\nm} TFLN plate with the mode displacement ($x$-component) shown in inset (i). By adding a silica underlayer, we instead excite an overtone $A_{21}$ resonance (ii, inset), which has weaker piezoelectric coupling (75$\times$ lower), but facilitates strong shear wave mediated AO interaction via Brillouin scattering at \qty{7.25}{\GHz} for MW-OT (b) By adding a silica lens on top of the LN, we can reduce the lateral wave leakage and increase the mechanical $Q_m$. In 3D, the lens reduces the overall cavity mode volume and increases the AO interaction strength $g$.}
\label{XBAR_basics}
\end{figure}

\kcb{To place our ideas in context, we start with a brief overview of XBAR modes, and how they need to be adapted to meet the requirements of MW-OT. Lateral field excited bulk acoustic wave resonators \cite{kadota20115, plessky2022xbar}, XBARs, were primarily investigated for wide bandwidth RF filtering applications, wherein it was realized that the anti-symmetric ($A_{1}$) Lamb wave has a very large piezoelectric coupling coefficient ($k^2_{eff}\approx$ 0.25-0.5) in $z$-cut thin-film lithium niobate with an in-plane lateral electric field applied along $y$.} 

\kcb{The interaction is mediated via the $e_{24}$ coefficient of LN, which is the largest at \qty{3.69}{\coulomb\per\meter\squared}. More importantly, for MW-OT applications, while the XBAR mode is a Lamb wave mode with a non-zero propagation constant along $y$, when operated near its cut-off frequency, it effectively serves as a shear mode thickness resonance, with a shear wave polarized along $y$, bouncing back and forth along $z$, while spatially confined between the electrodes. By adding a layer of silicon dioxide underneath, one can create an XBAR overtone resonator wherein one now excites a higher order $A_{n}$ resonance of the whole LN + oxide stack. It is important to note that even the overtone resonances are still mainly confined within the electrodes, which is critical for MW-OT operation.}

\begin{figure*}[htbp]
\includegraphics[width=\textwidth]{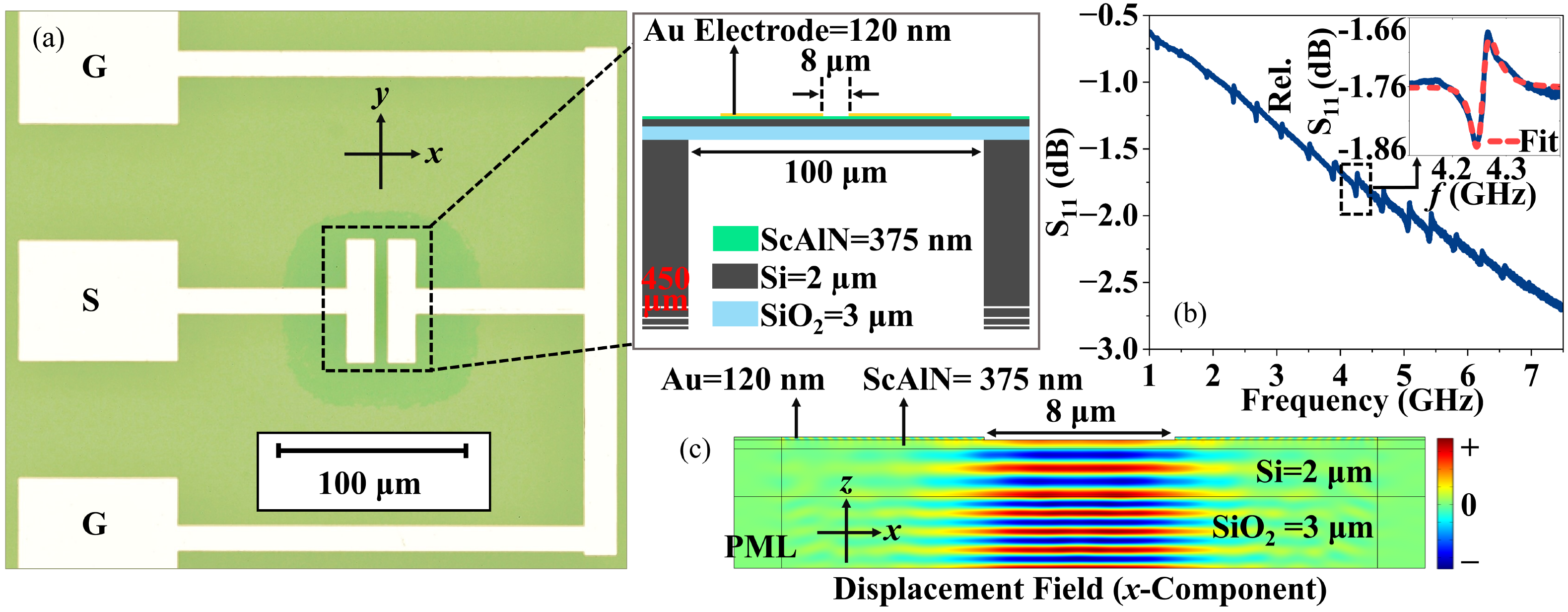}\\
\caption{(a) Microscope image of a representative XBAR device (flat-flat cavity without mode trapping) fabricated on a ScAlN-on SOI substrate. The layer thicknesses are noted in the inset. (b) RF reflection spectrum ($S_{11}$) of the device showing a series of successive overtone resonances of the cavity. A zoomed-in spectrum of one of the modes is shown in the inset showing an asymmetric (Fano) lineshape. The inset spectrum, labelled relative $S_{11}$, is corrected to remove the background tilt before fitting (dashed red) (c) 2D FEM simulation of one of the XBAR resonances under lateral field excitation. The $\vec{x}$-component of the displacement is plotted indicating the resonant mode has a shear vertical (SV) character, with the propagation direction along $\vec{z}$.}
\label{Unlensed_ScAlN_XBAR}
\end{figure*}

\begin{figure*}[tbph]
\centering
\includegraphics[width=\textwidth]{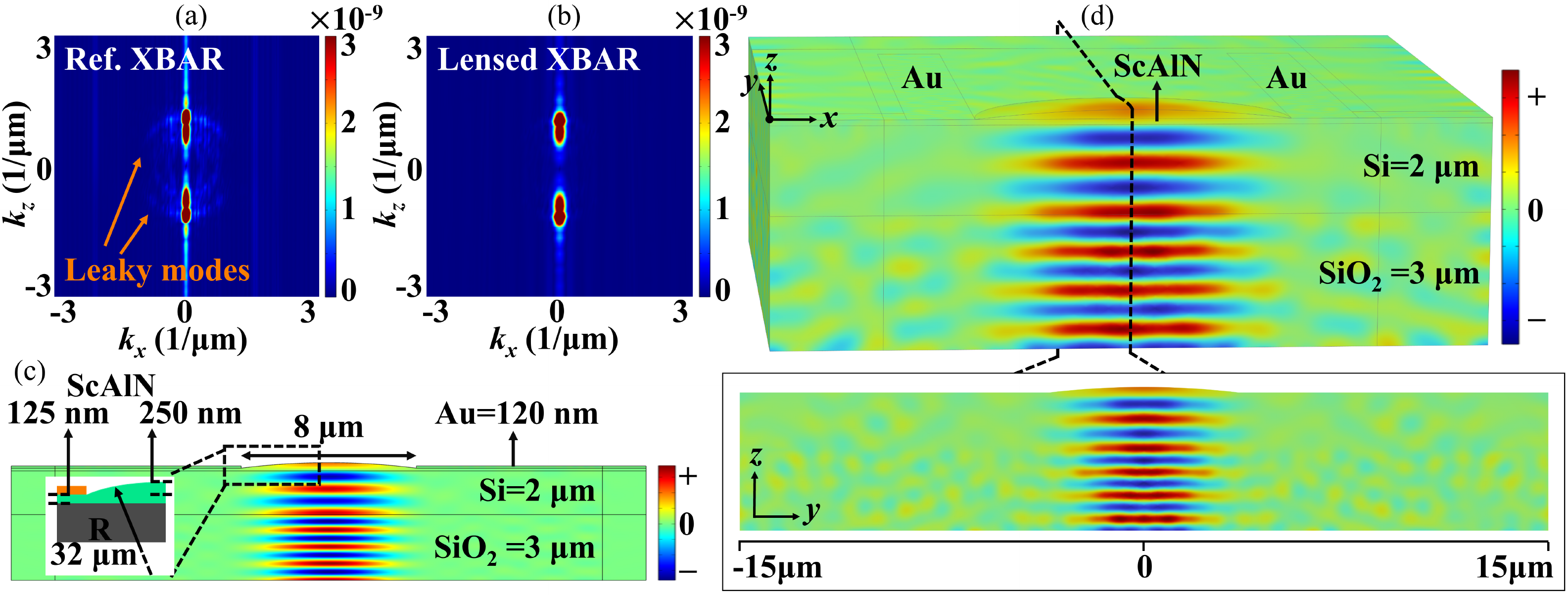}\\
\caption{Phonon trapping microcavity: By shaping the ScAlN layer into a spherical (lens-like) surface, one can prevent lateral leakage in the XBAR mode. (a) shows the $k$-space plot of the modal displacement of the XBAR cavity mode shown in Fig.\ref{Unlensed_ScAlN_XBAR}(c). Modes with non-zero $k_x$ which correspond to leakage through the sides appear as circular arcs as indicated. The same $k$-space plot for the lensed cavity mode (c) is plotted in (b) in which the modal leakage is minimized as shown by the absence of the non-zero $k_x$ modes. (d) 3D  modal confinement is evident from a full 3D FEM simulation of the lensed cavity showing the field ($x-z$ cross-section) is mainly confined under the lens at resonance. The inset shows the 2D cut along the orthogonal ($y-z$) direction indicated by the dashed line.}
\label{Diffraction_losses & lens_design}
\end{figure*}

\kcb{As discussed in eqn.\ref{eqn:SBS_g}, efficient AO interactions in bulk systems require operation at the Brillouin scattering frequency ($f_{BS}$). Assuming plane wave interactions, $f_{BS}=2nv/{\lambda}_o$, where $n$ is the refractive index at the optical wavelength $\lambda_o$ and $v$ is the acoustic velocity of the wave mediating the Brillouin interaction. The normal operating frequencies of current superconducting transmon qubits are in the 5-8 \qty{}{\GHz} range. Therefore, at telecom wavelengths ($\lambda_o$ = \qty{1.55}{\um}), achieving strong bulk wave Brillouin scattering at a frequency commensurate with current qubit technologies ($f_{BS}\approx$ \qty{7.25}{\GHz}) requires both low $n$ and $v$ \cite{valle2021cryogenic}, which is what silicon dioxide provides. We would like to note that the XBAR overtone resonance being a shear wave resonance is critical for ensuring a workable BS frequency compatible with qubits. The frequency compatibility is also why we chose not to engineer this interaction directly with a higher order Lamb wave $A_{n}$ mode of a thicker LN plate. While this would have potentially higher $k^2_{eff}$ and mechanical quality factor ($Q_m$), LN's higher $n$ pushes the shear wave $f_{BS}$ to $>$ \qty{10}{\GHz}.}

\kcb{$f_{BS}$ therefore sets the design frequency of our XBAR MW-OT. Starting with the $A_{1}$ resonance of a standard XBAR resonator with \qty{250}{\nm} thick LN, we adjust the LN thickness to \qty{350}{\nm} to achieve an $A_{21}$ overtone resonance with a \qty{5}{\um} silica underlayer. The thickness adjustment is needed to maximize $k^2_{eff}$ given the change in the strain ($S_{yz}$) distribution due to imperfect reflection at the LN-silica interface. $k^2_{eff}$ can be extracted from an FEM simulation by $k^2_{eff}=\frac{\pi}{2}\frac{f_r}{f_a}cot(\frac{\pi}{2}\frac{f_r}{f_a})$, with $f_r$ and $f_a$ the resonant and anti-resonant frequencies respectively \cite{hashimoto2009rf}. While the $k^2_{eff}$ is reduced by approximately 75$\times$ in going from the $A_1$ ($k^2_{eff}=0.32$) to the $A_{21}$ mode ($k^2_{eff}=0.0042$), it is still sufficiently high for efficient microwave to acoustic transduction, as we show below. We would like to note that MW-OT are expected to have operational bandwidths of $\approx$ \qty{1}{\MHz} \cite{weaver2025scalable}, which is much smaller than the $\approx$ \qty{500}{\MHz} bandwidth in 5G RF filters for which this mode was originally studied. The high $k^2_{eff}$ provides the headroom to adapt the XBAR design for MW-OT.}

\kcb{To build an efficient MW-OT using BS in an XBAR overtone resonator, two key problems must be addressed: the cavity mode volume needs to be reduced to increase $g$ as per eqn.\ref{eqn:SBS_g}, and the cavity mechanical quality factor $Q_m$ needs to be enhanced to ensure the optomechanical cooperativity ($C_{OM}\gg$ 1) is sufficiently high for quantum state transduction \cite{wu2020microwave}, as discussed below. One can achieve both these goals by structuring the XBAR resonator into a phonon trapping cavity by patterning an acoustic lens on the surface of the resonator.} 

\kcb{To understand phonon trapping, we note that the XBAR resonance is a Lamb wave mode with a small, but non-zero propagation constant. The major source of loss is therefore residual lateral mode leakage. In a simplified ray acoustics picture, the XBAR mode can be visualized as made up of a dominant shear wave propagation constant ($\vec{k_z}$), with a small $\vec{k_y}$ due to the Lamb wave origin. The wave therefore incurs a small off-axis propagation and it is this off-axis propagation that is refocused by the acoustic lens. The ratio ($|k_y|/|k_z|$) defines the lens curvature needed for efficient refocusing \cite{tanji2011interaction}.}

\kcb{An efficient MW-OT achieves high efficiency by matching the phonon injection rate into an optomechanical cavity with the AO scattering rate. As long as both rates exceed the (optical and mechanical) dissipation rates in the system, quantum state conversion from the microwave to the optical domain can be achieved \cite{wu2020microwave}. The rate matching can be expressed in terms of the electro-mechanical ($C_{EM}$) and optomechanical cooperativity ($C_{OM}$) following historical convention in cavity QED \cite{tanji2011interaction} as \cite{wu2020microwave}: $C_{EM}=1+C_{OM}$ with $C_{EM,OM}\gg$1. Here, $C_{EM} = \sqrt{k^2_{eff}}Q_m$ and $C_{OM} = \frac{4g^2N}{\kappa_o\gamma_m}$. Here $\kappa_o$ and $\gamma_m$ represent the cavity optical and mechanical decay rates respectively. As noted above, while the XBAR overtone configuration reduces $k^2_{eff}$, it is still sufficiently high to ensure $C_{EM}\gg$ 1 for expected $Q_m\approx$ \qty{5e3}{}. In traditional 1D optomechanical crystal approaches, the $k^2_{eff}$ for exciting the mechanical mode of interest is much smaller, and therefore places a far more stringent constraint on $Q_m$.}

\kcb{Given the obvious advantages of this approach on the acoustics front, it is important to consider the tradeoffs it presents on the optics front. The AO interaction \cite{balram2014moving} with the shear modes discussed here is mediated by $p_{44}\Re({E^*_yE_z})S_{yz}$, with $p_{44}$ the relevant photoelastic coefficient of (isotropic) silica, $E_{z,x}$ the optical field polarization and $S_{yz}=\partial{u_y}/\partial{z}$ is the relevant shear strain. The presence of the longitudinal $E_z$ field necessitates a small mode volume optical cavity and tight focusing of optical fields. The $p_{44}$ coefficient of silica is weaker than the $p_{12}$ coefficient traditionally involved in longitudinal wave Brillouin scattering. The reduction in $g$ is critical as $C_{OM}{\propto}g^2$ and needs to be carefully optimized in future designs.} 

\kcb{The other experimental issue that needs to be tackled is the alignment of a small mode volume, high $Q_o$ fiber cavity \cite{hunger2010fiber} around this XBAR overtone geometry. The high $Q_o$ is the main reason for pursuing this XBAR approach as it provides a pathway to achieving high $k^2_{eff}Q_m$ and high $Q_o$ simultaneously. While the existence of a high $Q_o$ optical mode around an XBAR overtone cavity can be confirmed numerically, the key challenge is to demonstrate that such a cavity can be assembled reliably and its alignment stability can be maintained. If we take experimentally demonstrated parameters ($Q_o, V_o$) for such fiber cavities and assume that similar functionality can be achieved here, $C_{OM}\gg$ 1 is within reach using this approach. In this work, as a first step towards this goal, we mainly focus on the acoustics, especially on the problem of trapping these shear mode overtone resonances in small mode volume devices.}

\section{Device Design and Fabrication}

\kcb{While LN represents the obvious platform for demonstrating these ideas, we demonstrate our ideas with scandium doped aluminum nitride (ScAlN) in-spite of its significantly weaker $e_{15}=e_{24}$ coefficient ($\approx$ 10$\times$ lower than $z-$cut LN). We chose this route because our main aim in this work was to demonstrate phonon trapping in XBARs as a route towards high $Q_m$ and the difficulty of etching LN reliably presents two key hurdles: presence of parasitic modes and insufficient suppression of lateral mode leakage. The lateral mode leakage is the more fundamental issue because it originates from the current difficulty of doing grayscale lithography and pattern transfer in LN using a physical etching process. While an oxide lens on top does help with lateral mode suppression, the ideal solution would be to shape the piezoelectric layer directly into a lens. This would suppress lateral mode generation \cite{tanji2011interaction} resulting in higher $Q_m$ and overall device efficiency. With this in mind, we proceed with ScAlN and tradeoff a significantly lower $k^2_{eff}$ in pursuit of higher $Q_m$. }

Figure \ref{Unlensed_ScAlN_XBAR}(a) shows a microscope image of a representative XBAR resonator device fabricated on a scandium aluminum nitride (ScAlN, with \qty{7}{\percent} Sc) on silicon-on-insulator (SOI) substrate with the respective layer thicknesses indicated in the inset. \kcb{The SOI device geometry was used here primarily because it was readily available for us to demonstrate the XBAR phonon trapping ideas, but the layer thicknesses were not optimized for MW-OT.}   

The membranes are fabricated (see Appendix A for details) by a deep reactive ion etching (DRIE Bosch process) of the silicon handle wafer. Laser lithography is then used to pattern electrodes aligned to the membrane as shown in the figure. This standard XBAR device serves as our reference, whose performance will serve as the baseline for the phonon trapping lensed cavity devices discussed below.

\begin{figure*}[htbp]
\centering
\includegraphics[width=\textwidth]{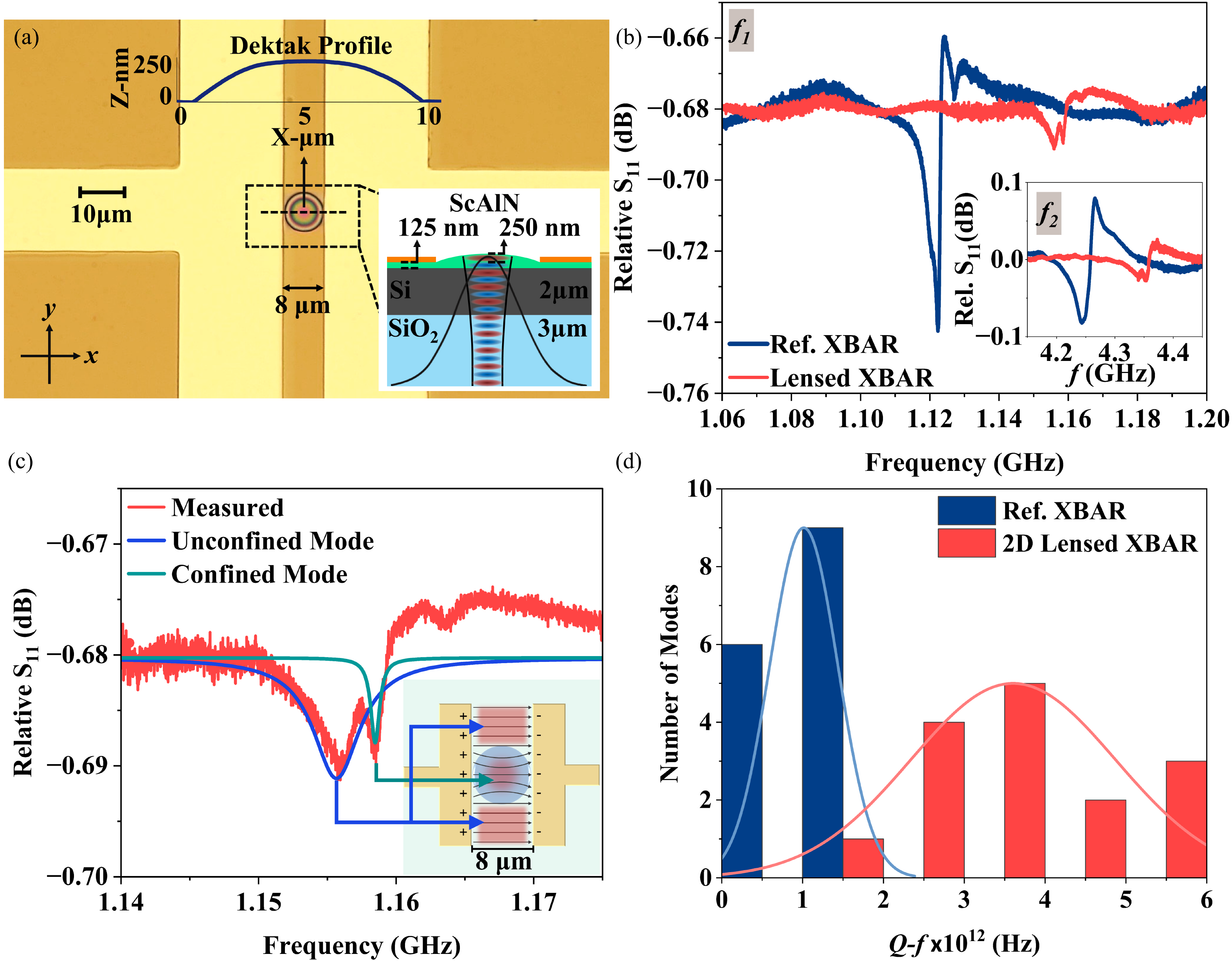}
\caption{(a) Microscope image of a representative lensed XBAR membrane device showing the spherical lens embedded in the ScAlN piezoelectric film. The cross-section of the device is shown in the inset with a Gaussian mode trapped in the cavity shown by illustration. The topography of the lens as measured by a profilometer (Dektak) is also shown via overlay. (b) RF reflection ($S_{11}$) spectrum for a representative mode in a lensed device (red) with the same mode in a reference (unlensed) device (blue) shown for comparison ($f_1$). A higher frequency comparison ($\approx$ \qty{4.3}{\giga\hertz}) is shown in the inset ($f_2$). The $S_{11}$ spectra are corrected to remove the background tilt before fitting, see Fig.\ref{Unlensed_ScAlN_XBAR}(b) and inset, and are indicated as Relative $S_{11}$. (c) Zoomed-in $S_{11}$ spectrum of the lensed device clearly showing a doublet with two modes with different $Q$. The lower $Q$ mode corresponds to the untrapped mode existing in the base ScAlN layer and the higher $Q$ mode is trapped in the lens region, as shown schematically in the inset. The $Q$ factors are extracted from a multi-peak Lorentzian fit to the tilt-corrected data. The individual Lorentzian fits from the two modes are shown in blue (unconfined, background mode) and green (confined, trapped mode) and the cumulative fit to the data in dashed black. (d) Histogram of the measured $fQ$ product of the modes from lensed vs reference devices showing that lensed devices give an overall 4$\times$ improvement in $Q$.  }
\label{Lensed_ScAlN_XBAR}
\end{figure*}

Figure \ref{Unlensed_ScAlN_XBAR}(b) shows the measured RF reflection ($S_{11}$) spectrum of the device in (a). A series of overtone resonances corresponding to successive Fabry-Perot resonances of the acoustic cavity are clearly visible. A zoomed-in inset into one of the modes is shown in the inset clearly showing that the measured resonance lineshape is asymmetric (Fano-like). The Fano lineshape arises due to the interference of the mechanical resonance lineshape (Lorentzian) with a background pathway which we currently attribute to the parasitic device capacitance. To fit the resonance, the background tilt in the response was removed, and we refer to such corrected spectra as relative $S_{11}$, see inset of Fig.\ref{Unlensed_ScAlN_XBAR}(b). Fitting this corrected spectrum with a Fano lineshape, shown by the dashed red curve, allows us to extract an effective mechanical quality factor ($Q_m$) of 365. In general, we find $Q_m$ in the range of 120-365. 

Figure \ref{Unlensed_ScAlN_XBAR}(c) shows a 2D finite element method (FEM) simulation of the device cross-section for one of the resonant modes at \qty{4.8}{\GHz}. As outlined above, the XBAR \kcb{overtone} resonances are \kcb{mainly confined in} between the metal \kcb{electrodes}. Fig. \ref{Unlensed_ScAlN_XBAR}(c)  shows that the mode is \st{a} \kcb{predominantly a} shear vertical (SV) bulk overtone resonance with the $\vec{x}$ component of the displacement indicated in the figure. The thickness resonance condition for the mode is satisfied along $\vec{z}$. \kcb{As discussed before, the Lamb wave nature of this mode results in a small $\vec{k}_x$ which contributes to leakage. We note that ScAlN films are isotropic in plane and therefore $\vec{x}=\vec{y}$. The lower $e_{15}$ coefficient of ScAlN, compared to LN, results in a very small $S_{11}$ dip at resonance, $\approx$ \qty{-0.25}{\decibel}.} 

The XBAR cavity mode in this reference geometry is only weakly confined between the electrodes ($x$-axis) as shown by the diffraction induced mode leakage in the FEM simulation Fig.\ref{Unlensed_ScAlN_XBAR}(c). Moreover, there is no geometrical confinement along the electrode length ($y$-axis) which makes this geometry unsuitable for building MW-OTs on account of the reduced $g$. By patterning the piezoelectric layer into a curved surface \cite{galliou2013extremely, kharel2018ultra}, we can spatially localize the mechanical mode and reduce the $A_{eff}$. In analogy with optics \cite{siegman1986lasers}, we effectively convert a flat-flat mirror Fabry-Perot cavity which is susceptible to modal diffraction into a curved-flat cavity wherein the spherical boundary effectively serves as a spherical mirror and re-focuses the acoustic field after a cavity round-trip.

To see the effect of the lens, it is helpful to look at the modal displacement in $k$-space, by taking the 2D Fourier transform of the mode displacement ($\vec{x}$)-component shown in Fig.\ref{Unlensed_ScAlN_XBAR}(c), shown in Fig.\ref{Diffraction_losses & lens_design}(a). In an ideal scenario, we would expect four discrete peaks at finite ${\pm}\vec{k}_z$ where $\vec{k}_z$ corresponds to $2\pi/\lambda_a$, where the $\lambda_a$ would be different in the silicon and silicon dioxide layer (ignoring the ScAlN contribution due to its thickness). As Fig.\ref{Diffraction_losses & lens_design}(a) shows, we do indeed see those 4 bright spots at finite $k_z$ and $k_x$=0, which are broadened to account for mode localization along $z$. In addition, we see faint circular arcs (indicated by orange arrows) at non-zero $\vec{k}_x$. These correspond to the residual leakage from the Lamb wave origin of the XBAR mode. We note that the finite electrode gap results in a broadening in $\vec{k}_x$ distribution, hence their appearance as circular arcs in $k$-space.  

By shaping the ScAlN layer into a spherical lens, as shown in Fig.\ref{Diffraction_losses & lens_design}(c), the same mode when visualized in $k$-space (Fig.\ref{Diffraction_losses & lens_design}(b)) shows a near-complete absence of the leaky modes (energy at non-zero $k_x$) indicating tight confinement and higher $Q$. For building MW-OT, we actually need modal confinement in 3D to maximize the AO interaction. Fig.\ref{Diffraction_losses & lens_design}(d) shows a 3D FEM simulation of the trapped lensed mode at resonance showing the energy is localized under the lens. We observe some excess lateral leakage in our 3D simulation compared to the 2D results and the trapped modes are not as pure as the ones we calculate via 2D FEM simulations. 

The two key geometrical parameters of the lens that we can control are the radius of curvature ($R$) and the lateral width ($w$) which sets the size of the spherical cap and determines the overall lensing and modal confinement. The size $w$ is mainly set by the inter-electrode gap, which is chosen as a compromise between two competing factors. Very small electrode gaps $<$ \qty{4}{\um} require tight focusing, high aspect ratio, small $R$ lens shapes which are difficult to fabricate. On the other hand, larger gaps both increase mode volume (lower $g$) and also reduce the overall $k^2_{eff}$ by lowering the device capacitance. Traditional XBAR devices get around this problem by using interdigitated electrodes that can help boost the capacitance, but this can not be applied directly to MW-OT because we are ultimately interested in co-localizing acoustic and optical fields. An electrode gap of \qty{8}{\um} is therefore chosen in our experiments. 

$R$ for the lens is determined mainly by our ability to faithfully transfer the spherical cap pattern from the photoresist to the piezoelectric ScAlN layer. As detailed in Appendix A, the spherical lenses are made by reflowing photoresist by baking which is known to produce a spherically smooth surface. The actual shape of the surface ($R$) can be controlled to an extent via the reflow temperature and time \cite{kirchner2019thermal}, and we find that the polymer surface is very close to a spherical cap in practice, as measured by a surface profilometer. This shape is then transferred to the ScAlN layer (cf. Fig.\ref{Lensed_ScAlN_XBAR}(a)) by reactive ion etching (RIE) and the shape of the lens is modified during this process from being a spherical cap to more of a flatter top spherical frustum according to the etch selectivity. Ensuring this pattern transfer maintains exact 3D shape fidelity is challenging due to the differential etching rates of the polymer and the ScAlN (etch selectivity 1.47), and the need to over-etch slightly to ensure the polymer is completely removed.

Figure \ref{Lensed_ScAlN_XBAR}(a) shows a representative fabricated phonon trapping XBAR device with a spherical lens embedded into the ScAlN layer. The total ScAlN device layer thickness is $\approx$ \qty{375}{\nm}, the lens is designed with an $R$ of \qty{32}{\um} and the lensed part is \qty{250}{\nm} thick. This leaves a residual \qty{125}{\nm} ScAlN base layer on which the electrodes sit as shown in the inset of Fig.\ref{Lensed_ScAlN_XBAR}(a). The shape of the lens as measured by a profilometer (Dektak) is shown via overlay in Fig.\ref{Lensed_ScAlN_XBAR}(a) clearly showing the flattened top after pattern transfer.  

\section{Characterization}

Figure \ref{Lensed_ScAlN_XBAR}(b) shows a representative RF reflection ($S_{11}$) spectrum of the lensed device (in red). A similar frequency mode from a reference (unlensed) device (as in Fig.\ref{Unlensed_ScAlN_XBAR}) is also shown for reference (in blue). The $S_{11}$ spectra are normalized to remove the background tilt (cf. Fig.\ref{Unlensed_ScAlN_XBAR}(b)). The bare device shows a larger $S_{11}$ dip on account of the larger volume of ScAlN (\qty{375}{\nm}). More interesting is the spectrum of the lensed device (in red). Fig.\ref{Lensed_ScAlN_XBAR}(c) shows a zoom-in to the mode which clearly shows a doublet with two modes with very different $Q$. The lower $Q$ mode arises due to the unconfined mode being excited in the base ScAlN layer ($t$=\qty{125}{\nm}), whereas the higher $Q$ mode is trapped in the central lens region. This is schematically illustrated in the inset of Fig.\ref{Lensed_ScAlN_XBAR}(c). We see these higher $Q$ modes across all frequencies, with one of the higher frequency (\qty{4.25}{\giga\hertz}) modes shown in the inset of Fig.\ref{Lensed_ScAlN_XBAR}(b). The $Q$ factors are extracted by performing a multi-peak Lorentzian fit to the data and the exact $Q$ value is sensitive to the signal level, which is limited in our experiments by the small $S_{11}$ dip magnitudes. On the other hand, as the same fitting procedure is applied to extract the $Q$ of the background mode, the ratio between the $Q$ of the confined and unconfined modes is a more robust indicator of phonon trapping.

\begin{figure}[htbp]
\centering
\includegraphics[width=0.5\textwidth]{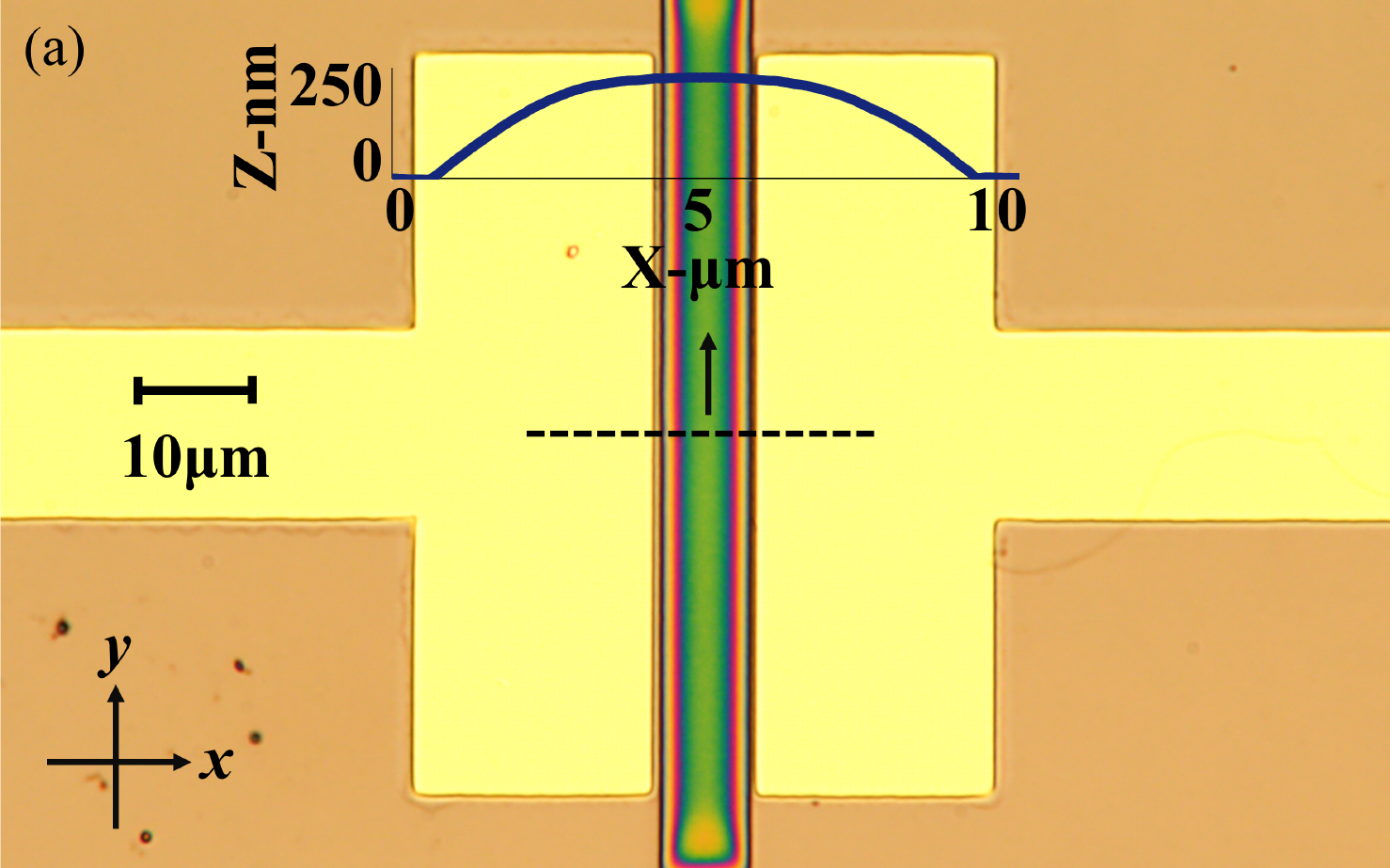}\vspace{2mm}
\includegraphics[width=0.5\textwidth]{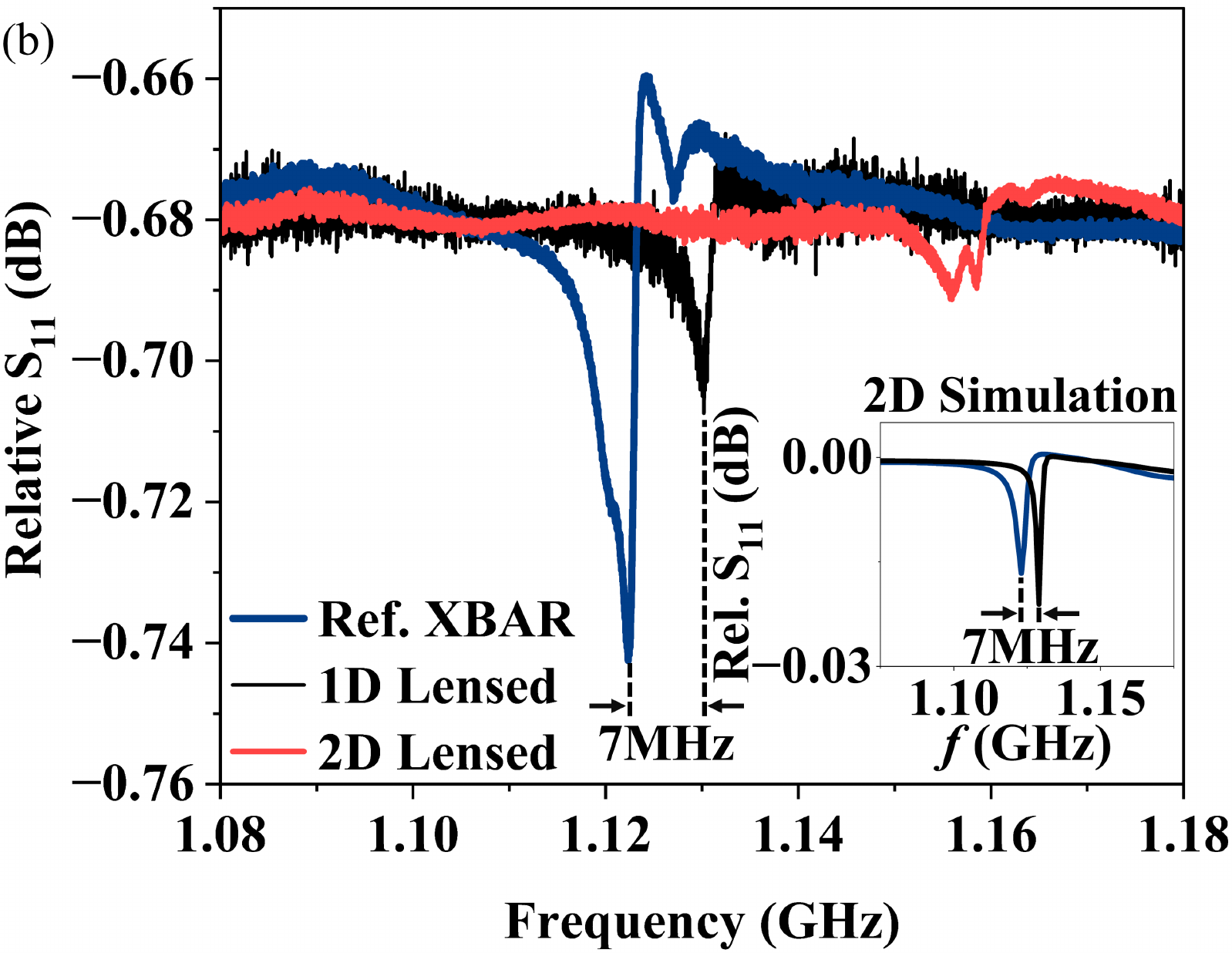}\\
\caption{Microscope image of a 1D lensed XBAR device mode-trapping cavity fabricated on ScAlN using a reflow etching process, along with a Dektak scan showing a lens thickness of 250~nm and a radius of 32~\textmu m, placed between electrodes with a gap of \qty{8}{\um}. (b) RF reflection ($S_{11}$) comparison between bare, 1D-lens and 2D-lens devices. The shift in frequency between the bare and the 1D lens devices agrees well with the 2D FEM simulation (inset) showing the geometry works as intended. The 2D lens data has additional structure due to the presence of the background mode (cf. Fig.\ref{Lensed_ScAlN_XBAR}(c)).}
\label{1D Lens}
\end{figure}

We can quantify this $Q$ improvement by plotting a histogram of the $fQ$ product for both lensed and reference devices, wherein we take the higher $Q$ mode to represent the trapped (lensed) mode. Such a histogram is shown in Fig.\ref{Lensed_ScAlN_XBAR}(d) which clearly shows that adding a spherical lens helps improve $Q$ by $\approx$ 4$\times$. While the $Q$ improvement is a promising signature, it does not directly point to spatial mode confinement (phonon trapping) under the lens, which is also critical for MW-OTs. Ideally, one would like to locally map the displacement at the two frequencies (corresponding to the unconfined and lensed modes) and show that the spatial mode confinement is correlated with $Q$. We attempted the spatial mode mapping, but were unable to resolve the acoustic displacement in these XBAR devices, due to a combination of low $k^2_{eff}$, and the shear mode nature which places the predominant displacement along $\vec{x}$ instead of the out-of-plane component $\vec{z}$, which is what the experiment is sensitive to. From simulation, the surface $\vec{z}$ is lower than $\vec{x}$ by $\approx$ 40$\times$ (2D) - 100$\times$ (3D)  for these devices. 

In the absence of direct mode mapping, we turn to control experiments to infer mode localization. Figure \ref{1D Lens}(a) shows a 1D variant of the lens where the mode confinement is engineered along only one axis ($\vec{x}$), between the electrodes. Figure \ref{1D Lens}(b) shows the $S_{11}$ spectrum of such a 1D-lens device. We overlay the spectra of the reference (unetched) XBAR device and the 2D lens device as well. We can see that the $S_{11}$ dip of the 1D-lens device is lower than the bare device on account of reduced piezoelectric material volume. The bare device has a uniform film of \qty{375}{\nm} thickness, whereas the 1D-lens is shaped like the inset of Fig.\ref{Lensed_ScAlN_XBAR}(a). More importantly, the frequency shift between the reference and 1D-lens device (\qty{7}{\mega\hertz}), tracking modes with the same number of nodes along the propagation direction $\vec{z}$, agrees reasonably well with 2D FEM results (shown in the inset). This proves, at least for 1D confinement, the phonon trapping works as intended.

The 2D lens is further shifted compared to the 1D lens device and the spectrum is more complicated. This is because of the presence of the background XBAR modes in the \qty{125}{\nm} ScAlN layer, as shown by the red shaded regions in the inset of Fig.\ref{Lensed_ScAlN_XBAR}(c). Making an unambiguous frequency shift comparison with FEM, as before in the 1D lens case, is more challenging here because of numerical issues related to mesh size and the presence of additional modes in the 3D volume. Nevertheless, the frequency trend from unconfined to 1D to 2D confinement is correct and the $Q$ factor increases as expected, so we believe this provides evidence for acoustic mode trapping in these lensed XBAR devices.

\begin{figure*}[!t]
\centering
\includegraphics[width=0.96\textwidth]{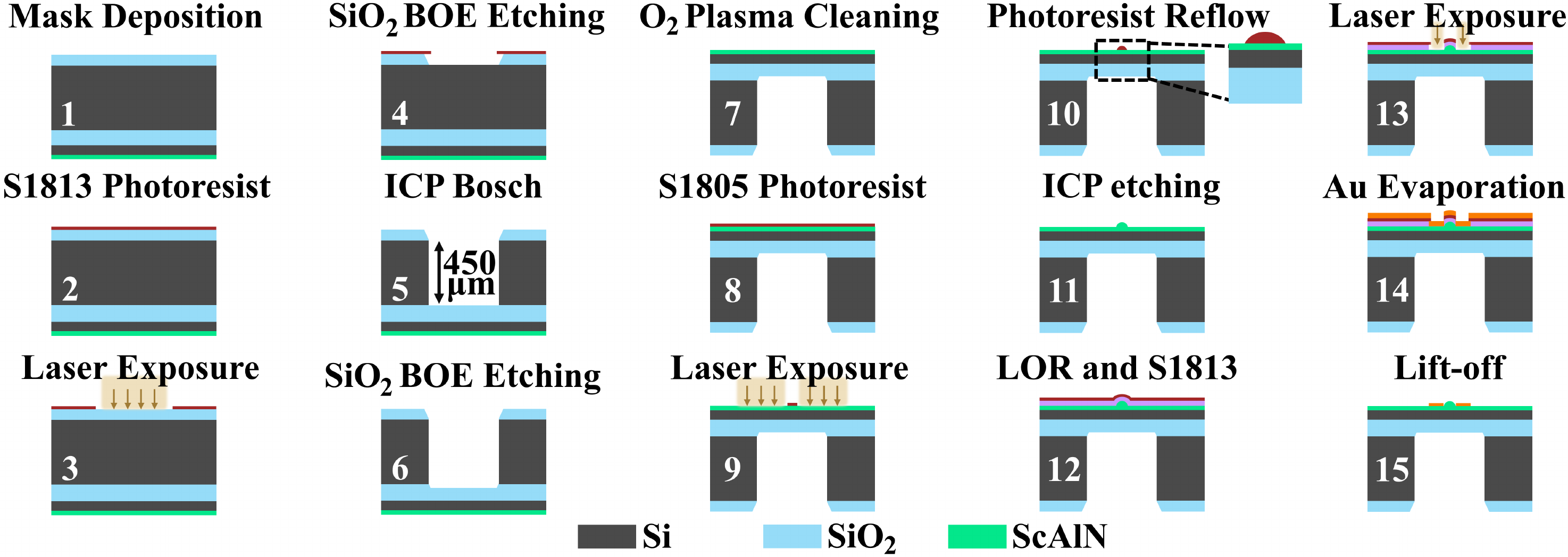}\\
\caption{Schematic illustration of the lensed XBAR fabrication process. The process includes backside patterning and substrate etching using a SiO$_2$ hard mask, membrane release via deep reactive ion etching  (ICP Bosch), and subsequent lens fabrication on a 375~nm-thick ScAlN layer using photoresist reflow and plasma etching. 1D and 2D lenses are formed with final thicknesses of $\sim$250~nm and radii of $\sim$32~\textmu m. This is followed by standard lithography steps for electrode fabrication, including gold evaporation and lift-off. S1805 and S1813: photoresists, LOR: lift-off resist, BOE: buffered oxide etch.}
\label{Fab Process}
\end{figure*}

\section{Conclusions}

While we have shown small mode volume XBAR resonators in this work, a few issues need to be addressed before these devices can operate as efficient MW-OTs. \kcb{As discussed above, we worked with ScAlN here mainly because of the possibility of structuring the piezoelectric in the pursuit of high $Q_m$. But useful MW-OT using this approach will require LN as outlined earlier to ensure sufficient $k^2_{eff}$ in an overtone geometry. While there has been immense progress in LN dry etching, gray-scale lithography and 3D shape transfer are still not available as standard. This means that near term designs will have to rely on an oxide lens as outlined in Section 2 with its associated $Q_m$ limitations. Suppression of parasitic modes in this geometry is also an important open question.} 

The second issue relates to improvement in $Q$. While we have clearly shown an improvement in $Q$ over the reference XBAR mode, the highest $Q$ we observe in our devices are $\approx$ 2045 and our $fQ$ products overall are $<$ \qty{6e12}{} which is lower than both bulk HBAR and integrated phononic devices \cite{gokhale2020epitaxial, bicer2023low}. In contrast to traditional HBAR and FBAR devices, as the acoustic field overlap with the metal is minimized, XBAR devices should in principle have lower acoustic dissipation,  Understanding the residual sources of loss and leakage, especially from 3D FEM simulations, and working out the optimal lens shape given fabrication constraints is a key element to improving overall device performance. As noted above, since MW-OTs are expected to have operating bandwidths around \qty{1}{\MHz}, improvements in $Q$ can help mitigate a lower overall $k^2_{eff}$.

Finally, while we have focused purely on the mechanical aspects of MW-OTs in this work and have been (mechanical) frequency agnostic in some sense, as noted above, MW-OTs in this XBAR-like geometry can only achieve high AO coupling strengths ($g_0$) while operating at the Brillouin frequency \cite{renninger2018bulk, valle2019high, valle2021cryogenic}. To achieve an interaction at \qty{7.25}{\GHz} \cite{valle2021cryogenic} to be compatible with current qubit frequencies, BS needs to be engineered in the silica layer. Given the importance of mechanical dissipation in MW-OT \cite{wu2020microwave}, the $Q_m$ that can be obtained in SiO\textsubscript{2} at low temperatures at \qty{}{\GHz} frequencies will be critical to achieving high conversion efficiencies, and this remains an open question, although there are promising low frequency results in fused silica \cite{schroeter2007mechanical, numata2002intrinsic} and quartz \cite{galliou2013extremely} that indicate low mechanical dissipation at cryogenic temperatures.

\section*{Acknowledgments}
We would like to thank Stefano Valle and Vinita Mittal for early work in this area, and Mahmut Bicer and Haoyang Li for feedback on the manuscript. This work was supported by the European Research Council (758843), the U.K.'s Engineering and Physical Sciences Research Council (EP/T517872/1, EP/N015126/1, EP/W035359/1) and the UKRI Frontier Research Guarantee (EP/Z000688/1). The ScAlN-on-SOI wafers were acquired from Tyndall Institute as part of an EU-Ascent+ program (871130, 654384), and we thank Veda Sandeep Nagaraja for guidance. ES would like to acknowledge support from the University of Bristol's EPSRC Impact Acceleration Account (IAA).

\section*{Appendix A: Device Fabrication}

Figure~\ref{Fab Process} illustrates the fabrication steps for the phonon-trapping lateral-field-excited suspended XBAR. As ScAlN is susceptible to standard silicon wet etchants (like potassium hydroxide (KOH) and tetramethyl ammonium hydroxide (TMAH)), during membrane release, a deep reactive ion etch (DRIE) was used to punch through the silicon substrate to release the membrane. 

To etch 450~\textmu m of silicon, a 5~\textmu m-thick SiO$_2$ layer was deposited on the Si substrate using plasma enhanced chemical vapor deposition (PECVD) as a hard mask. A square window for the DRIE etch was patterned by etching the SiO$_2$ window using hydrofluoric acid (BOE). After the DRIE etch, a BOE etch was applied to etch the plasma damaged oxide on the backside of the membrane. Overall, with the DRIE and the BOE etch, we remove \qty{3}{\um} of oxide. Finally, the sample was examined and cleaned with O$_2$ plasma prior to membrane release.

Following membrane release, 1D and 2D lenses were fabricated on the \qty{375}{\nm}-thick ScAlN layer, with lens thicknesses of approximately \qty{250}{\nm} and radii ($R$) of $\approx$32~\textmu m. The remaining ScAlN thickness beneath the lenses was around \qty{125}{\nm}, as shown in the inset of Fig.\ref{Lensed_ScAlN_XBAR}(a). Lens fabrication employed a photoresist reflow process: the photoresist was initially patterned into cylindrical shapes, then reflowed into hemispherical forms driven by polymer chain mobility and surface tension \cite{kirchner2019thermal}. Once the desired lens shapes were obtained, plasma etching was used to remove the photoresist and transfer the lens pattern to the substrate, with a selectivity of approximately 1.47.

The lens radius of approximately 32~\textmu m is chosen based on practical considerations in transferring the reflowed photoresist pattern onto the ScAlN layer while preserving a smooth, nearly spherical surface. Control over the radius ($R$) is achieved in part by adjusting the photoresist reflow temperature and duration \cite{kirchner2019thermal}. Profilometer measurements confirm that the final lens shape closely resembles a spherical cap, with a slight flattening at the top. Finally, Cr/Au electrodes are patterned around the lens using another aligned lithography step.

%\nocite{*}
%\clearpage

\bibliography{References}

\end{document}